\documentclass[runningheads]{llncs}
\usepackage[utf8]{inputenc}
\usepackage[english]{babel}
\usepackage[T1]{fontenc}
\usepackage{comment}
\usepackage{graphicx}
\usepackage[hidelinks]{hyperref}

\begin{document}

\title{Flexible Educational Software Architecture\thanks{This work was supported by the German Federal Ministry of Education and Research for the tech4comp project under grant No 16DHB2102.
}}
\subtitle{at the example of EAs.LiT v2}
\titlerunning{Architecture of EAs.LiT v2}

\author{Roy Meissner\inst{1}\orcidID{0000-0003-4193-8209} \and Andreas Thor\inst{2}\orcidID{0000-0003-2575-2893}}

\institute{Leipzig University, Faculty for Education, Leipzig, Germany
\email{roy.meissner@uni-leipzig.de}\and
Leipzig University of Applied Sciences (HTWK Leipzig), Leipzig, Germany
\email{andreas.thor@htwk-leipzig.de}}

\maketitle

\begin{abstract}
EAs.LiT is an e-assessment management and analysis software for which contextual requirements and usage scenarios changed over time. Based on these factors and further development activities, the decision was made to adopt a microservice architecture for EAs.LiT version 2 in order to increase its flexibility to adapt to new and changed circumstances. This architectural style and a few adopted technologies, like RDF as a data format, enabled an eased implementation of various use cases. Thus we consider the microservice architecture productive and recommend it for usage in other educational projects. The specific architecture of EAs.LiT version 2 is presented within this article, targeting to enable other educational projects to adopt it by using our work as a foundation or template.

\end{abstract}

\section{Introduction}
EAs.LiT is a web application handling e-assessment item and learning outcome lifecycle, - management, - annotation, as well as automated exam creation and exam analysis, as introduced by Thor et al. in \cite{Thor2017easlit}. In 2018 the German Federal Ministry of Education and Research granted further development of the software EAs.LiT as part of the tech4comp project.
With respect to the demands of this research project, but also as time passed by, new and adapted requirements and usage scenarios arose for EAs.LiT. The original software was built using a monolithic system architecture and imposed several challenges regarding maintenance, expandability, reusability, and repurposibility as of the evolved context, like argued by Newman in \cite{newman2015building} for monolithic software systems in general. Thus we decided to develop EAs.LiT version 2 with a different and more suitable software architecture, namely the microservice architecture, which tackles the named topics. It offers even more advantages in comparison to a monolithic architecture, like argued in detail by Newman in \cite{newman2015building}.

We consider the microservice architecture productive and recommend it for usage in other educational projects. Thus the focus of this paper is to present the architecture of EAs.LiT version 2 in section \ref{easlitArch}, as well as its accompanying technologies RDF and JSON-LD. Furthermore we showcase in section \ref{useCases} the implementation of specific use cases, which were eased by the architecture and used technologies. Our main goal is to enable educational projects to adopt the architectural style, using our work as a template. To achieve this goal, we additionally present in section \ref{relwo} related work that focuses on software projects which also use a microservice architecture. We end by summarizing our approach and emphasizing unique features, as well as presenting future development topics.

\section{Related Work}
\label{relwo}

SlideWiki, which may be considered a Github for slides in the educational context, uses a microservice architecture, which was presented by Khalili et al. in \cite{slidewiki2018architecture}. They mainly focus on distributed development and collaboration of teams. In contrast to their work and software we do not focus much the social aspects of the architecture, but on technological aspects, like the use of semantic technologies in all architectural layers. Furthermore we showcase novel integration scenarios, like the use of spreadsheets as a batch processing frontend.

Kumar et al. propose in \cite{Kumar2017microservice} to utilise a microservice architecture for e-learning systems. They unveil that it is a solution to several downsides of current monolithic e-learning systems and constructed a simple, but theoretical example. Similar to our own work, they propose to use several web technologies, but are not providing an implemented proof of concept and only scratch the topics surface.

Fazakas et al. present in \cite{Fazakas2018learningtools} a novel virtual classware technology which uses a microservices architecture. They mainly focus on their use case and developed software and only lightly touch the software architecture topic. Microservices are named an enabling technology for their solution and they used similar web technologies to the ones we used. In contrast we focus less on the built software, but much more on the architecture itself and why using it enlarges the flexibility of the software system.

Rad et al. propose in \cite{Rad2017edunet} an "innovative technology that implements problem-based learning and projects-based learning concepts". They utilise a cloud infrastructure for their solution and run several microservices as part of it. No exploration of the decision was presented and they seem to use it as it fits their needs and is supported by the utilised cloud stack.

There are much more software architecture publications available, focusing on microservices. In contrast to the presented literature these publications focus on different aspects of the architecture, like scalability and incremental updates, target different use cases, like delivering large amounts of data to end users, and are not positioned in the educational software context. Thus we do not present these here, but wanted to mention their existence.

\section{Architecture of EAs.LiT version 2}
\label{easlitArch}

\begin{figure}[ht]
    \centering
    \includegraphics[width=0.9\linewidth]{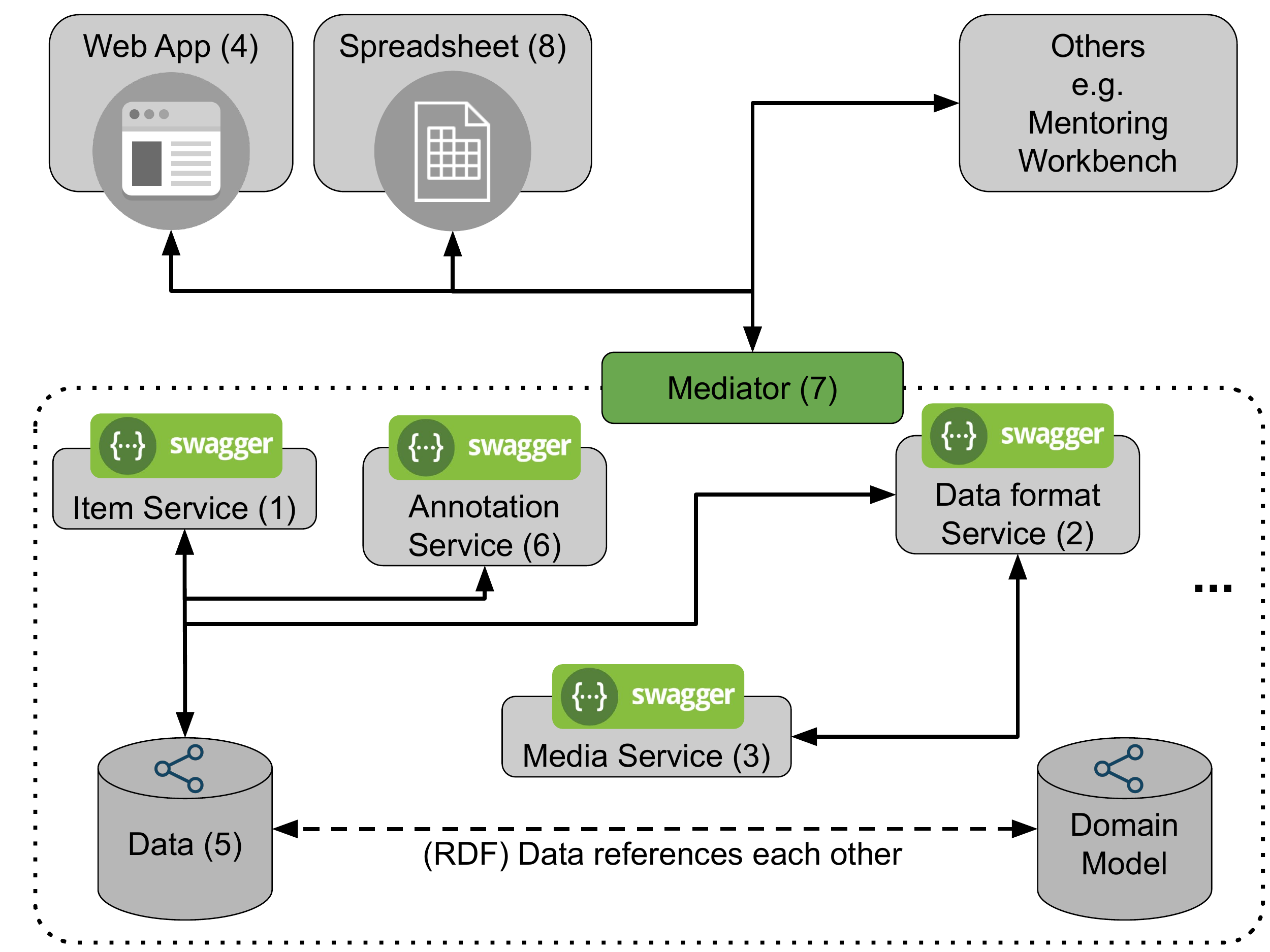}
    \caption{Bird's-eye view of the EAs.LiT version 2 architecture}
    \label{fig:architecture}
\end{figure}

EAs.LiT version 2 implements a microservice architecture, as introduced and argued in detail by Newman and Wolff in \cite{newman2015building,wolff2017microservices}. This means that the software system is not executed as a monolithic system, like EAs.LiT version 1, but is composed of several independent and fine grained services. These services are built using a domain-driven design approach, described by Evans in \cite{evans2004domain}. Each service implements a distinct and enclosed domain of the whole system. As of the current project progress there are services for the domains: 1) item management, 2) data format conversion and data import/export, 3) media management, 4) the end user frontend, 5) data management (database), 6) item annotation, 7) traffic management (a reverse proxy), and 8) item batch editing. Within the following paragraphs we present and discuss all parts of the new EAs.LiTs' architecture, using a birds'-eye view depicted with figure \ref{fig:architecture}.

\vspace{-1.2em}
\vspace{-1.5em}

Figure \ref{fig:architecture} shows in the upper left corner two end user applications: the main frontend of EAs.LiT (4), as well as a spreadsheet based item batch editing interface (8). These two services share their end user focus and usage of other systems services, but are built using different technologies and targeting different use cases. 4 is built as a browser application and allows usage of EAs.LiT by non technical user groups, i.e. by educators, by providing an easy to use and modern user interface. To get an impression of the software we refer to figure \ref{fig:frontend2} from section \ref{workbench}. As EAs.LiTs' main functionality is intended for educators and psychologists, the frontend is a central part of the whole project and is used for service composition. Javascript is used as a programming language and it was built with a modern NodeJS stack, drawing upon Vue.js, React, Bootstrap and Webpack. Furthermore the frontend utilises a Flux architecture and JSON-LD, which we argue about below. 8 is also an end user frontend, but in contrast to 4 it is a spreadsheets' plugin, built for batch editing items. We explain this special frontend and use-case in more detail in section \ref{excel}.

Service 7 (traffic management) is a reverse-proxy server and used to route incoming and outgoing traffic correctly and to handle traffic encryption, rate limiting, authentication and other related topics. It is a mediator (like named in figure \ref{fig:architecture}) or gatekeeper between software running on our infrastructure and software executed elsewhere. Its purpose is to free service developers from implementing the just named topics to each service and allowing them to focus on their specific domain only. Currently we use a NGINX based reverse-proxy server\footnote{Reverse-Proxy: \url{https://hub.docker.com/r/jwilder/nginx-proxy}}.

Microservices 1,2,3 and 6 focus on different distinct domains, like item management and item annotation. All of these services implement a representational state transfer (REST) interface, utilizing OpenAPI (formerly known as swagger) for standardized interface descriptions. A REST interface allows to easily incorporate services from each other, from the official fronted (4), from specialized frontends (like 8) and also from third parties, like presented in section \ref{workbench}. Apart of the two named shared technologies, the various services are built using different frameworks and different programming languages. This is a general aspect of the microservice architecture and allows to utilise the various specialized features of different frameworks, programming languages and also special abilities of programmers, like argued about in \cite{newman2015building}. For instance the import/export service is built using Java, because of its stream processing and interface capabilities. The annotation service is built using Python, as the specialized framework scikit-learn is available for this language. The item management-service is implemented using Javascript because of its native JSON interface and developer skills.

Lastly we depicted various data storage services (5) in figure \ref{fig:architecture}. We currently use two graph databases for storing data from different services. In general the microservice architecture allows to use different data storage engines for different services and use-cases, e.g. a relational database where it is appropriate. We have incorporated this possibility to separate the data of EAs.LiT from the domain specific knowledge graphs, which are introduced in section \ref{knowledgeMap}. Thus we use different database instances, but all are graph stores as these fit our use cases best.

As of noticeable technologies, RDF \footnote{RDF: \url{https://www.w3.org/TR/rdf11-primer/}}, serialized to JSON-LD\footnote{JSON-LD: \url{https://www.w3.org/TR/2019/CR-json-ld11-20191212/}}, is used as the central data format through all of the above described services. These technologies impose extra effort for data management and processing, but ease a lot of use-cases on different levels, like presented in section \ref{knowledgeMap}. RDF is a open world data format and thus easily extended with new data, without running into structural limits like introduced by entity relationship models. JSON-LD, which is one possible serialization of RDF data, adds a context to traditional JSON documents. This additional context may be utilised by software like the frontend (4), but may also be ignored and thus allows to treat the data as traditional JSON data, like done by the import/export service (2). Furthermore JSON-LD allows to decouple services on the data level itself, as traditional JSON identifiers are optional and may easily be exchanged by using a different context, as well as to restructure data as of the data input process, like shown by Veltens in \url{https://blog.codecentric.de/2018/07/datenaustausch-mit-json-ld/}. We are not only using RDF data, but Linked-Data\footnote{Linked-Data: \url{https://www.w3.org/wiki/LinkedData}}, which allows to decouple the executed frontend from whole EAs.LiT instances, as received data identifies its origin.

\section{Enabled Use-Cases}
\label{useCases}
Within the following subsections we present special use cases and implementations of these, which the described architecture from section \ref{easlitArch} enabled and eased to implement. We treat these use cases as proof of concepts, research prototypes and examples on how the named technologies may be efficiently applied.

\subsection{Mentoring Workbench Integration}
\label{workbench}
The Mentoring Workbench (MW) is the main software product of the tech4comp project (to be described in future work) and provides several web-components, that are activatable in different places of a Learning Management System (LMS). Some components are usable by students, others are usable by educators and supervisors. One component the MW provides is a personal exam analysis tool for students. It links available e-assessment items to exam results from the LMS and pulls additional data from EAs.LiT to create a detailed analysis of the individual exam results. The component relies on item annotations, such as levels of Blooms taxonomy about the cognitive learning domain \cite{bloom1973taxonomie}, manually added by educators. The annotated items are pulled and processed by the MW and merged with the LMS data in order to e.g. inform students about reached performance levels.

The architecture presented in section \ref{easlitArch} eased this process as the MW pulls in items via the already existing item management API (1). It might not be obvious, but this API has naturally evolved because of the microservice architecture and the applied separation of domains. The frontend and the item management are two separate domains, thus realised as different services, which communicate through a REST API. So an API for item creation, update, removal and retrieval already exists and is reused by the MW. In comparison to EAs.LiT version 1, either a manual data export, a new API, or direct database access for third party software needs to be implemented to achieve the same functionality.

\subsection{Spreadsheet based Batch Processing}
\label{excel}
\begin{figure}[hb]
    \centering
    \includegraphics[width=\linewidth]{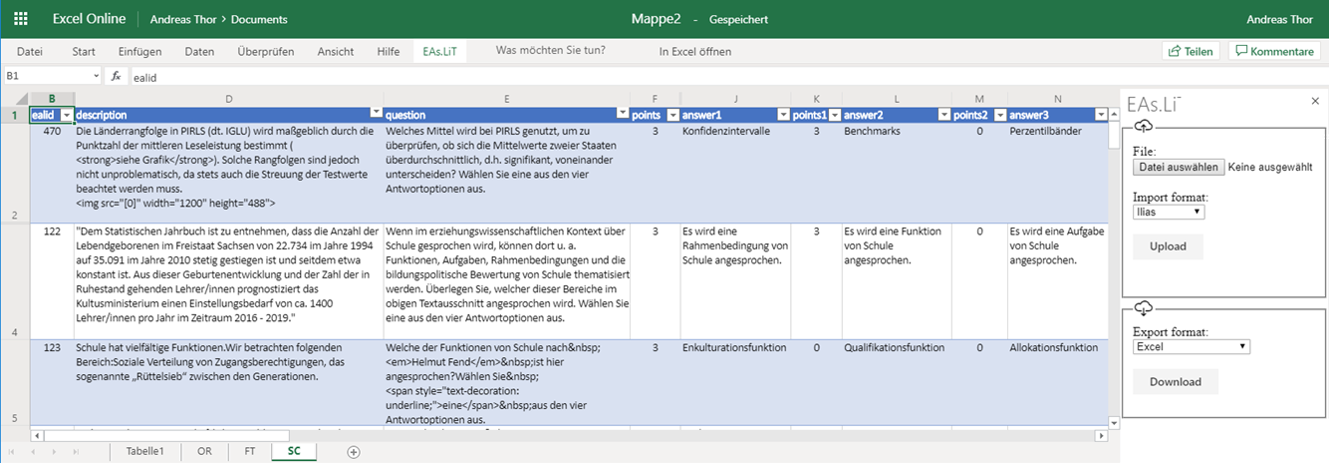}
    \caption{Example of the batch processing interface}
    \label{fig:excel}
\end{figure}

As broached in section \ref{easlitArch} we implemented an item batch processing interface as a plugin for Microsoft Excel, depicted in figure \ref{fig:excel}. This plugin is written in Visual Basic for Applications and automatically included as item sets are exported from EAs.LiT via the import/export service or via the frontend. Nevertheless the plugin is also usable on its own when provided with an EAs.LiT instance reference (URLs for the item management and import/export services). The plugin allows to fetch item sets from EAs.LiT, to use the capabilities of spreadsheets for batch editing and processing and to upload the updated items back to their origin. Spreadsheets efficiently allow various batch editing capabilities for items, like to increment the number of given points, to compare and exchange answer texts, to revise item annotations, like presented in section \ref{workbench}, and to create a lot of similar items by copy and paste operations. In comparison to EAs.LiT version 1, several manual steps for data export and import via the browser interface would have been needed. Even more importantly the already existing services supported this use case without the need to apply any changes to their API.

\vspace{-1em}
\vspace{-2em}

\subsection{Domain Model Integration}
\label{knowledgeMap}

\noindent One goal of the tech4comp project is to create preferably complete knowledge graphs for specific domains, like the educational discipline. For various use cases it is desirable to link learning outcomes and items to this graph, e.g. to provide the ability to analyse the item/domain distribution. As broached in section \ref{easlitArch} items and learning outcomes managed by EAs.LiT are stored as RDF data, as is the knowledge graph. We hinted in section \ref{easlitArch} that RDF allows to link data sets, but does not require to merge these in order to do so. So for example a specific item is annotated with a specific domain from the knowledge graph by adding the URI of this domain to the item. Thus the used data format allows a fast and straight forward implementation of this use case. Because RDF is an open-world data format, adding new data (even on a structural level) will not break any existing services, like the item management service (1). It even enables to utilise this new data, i.e. to offer new abilities to API users and eventually frontend users.

\begin{figure}[ht]
    \centering
    \includegraphics[width=\linewidth]{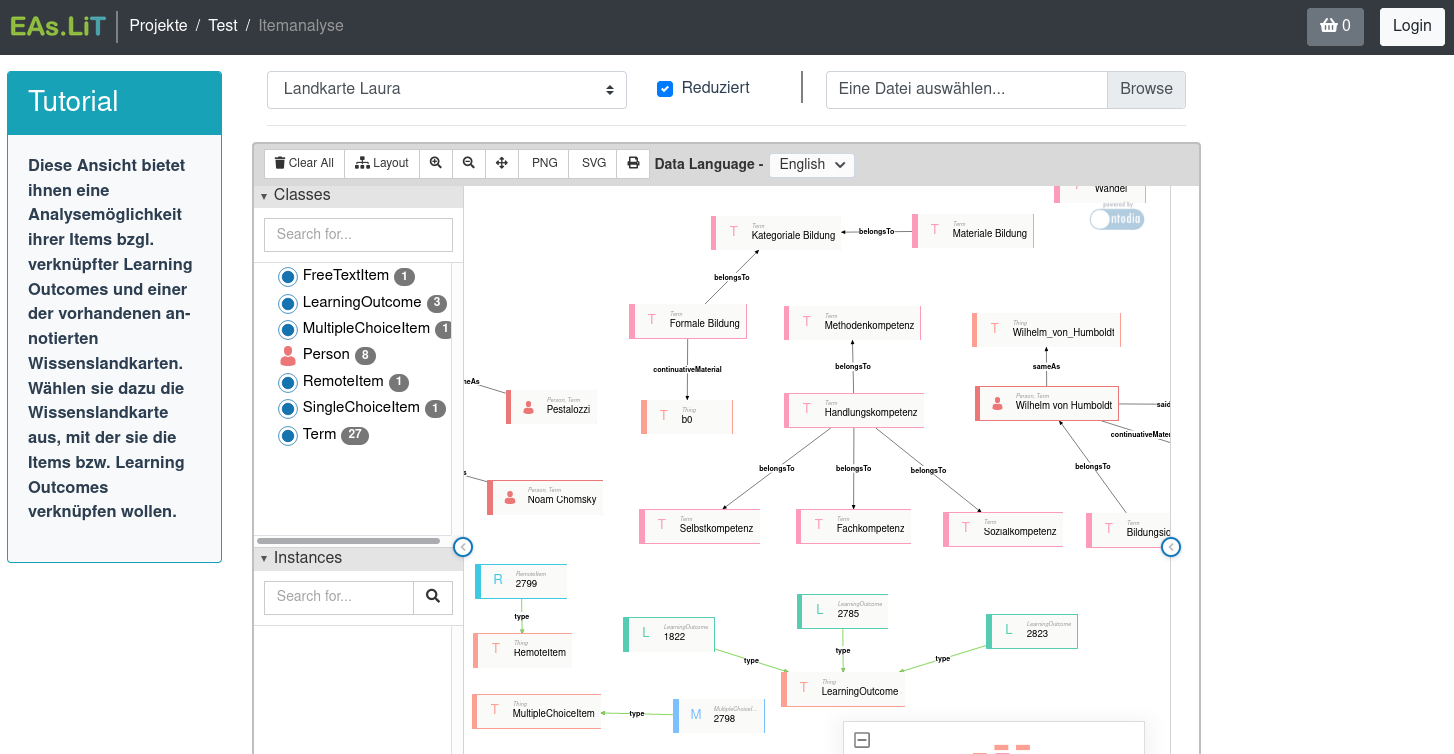}
    \caption{EAs.LiTs item/domain distribution analysis interface}
    \label{fig:frontend2}
    \vspace{-1.5em}
\end{figure}

Based on these facts we realised a frontend component which allows to visually analyse the item/domain distribution, depicted in figure \ref{fig:frontend2}. This component merges the received item and learning outcome data, as well as a reduced view of a knowledge graph, received directly from the knowledge graph store. The result is presented as an explorable canvas to end users. Because of the RDF technology, various RDF data serializations and the microservice architecture, we were able to skip implementing complex merge strategies or rule sets as the different data sets just need to be appended into a single data record prior to being displayed.

\section{Summary \& Future Work}

A particular adoption of the microservice architecture for an educational software project has been presented and explained at the example of EAs.LiT version 2. We focused on the concrete software architecture in order to enable third party projects to learn about the architectural advantages and to adapt it for their own software projects. Furthermore novel use-cases have been discussed, which were either enabled or at least eased to realise because of the used architecture. Several technologies are accompanying the various use cases of EAs.LiT and thus we presented RDF and JSON-LD as key data formats, used through all the architectural layers. Characteristics of these two technologies complement benefits of the microservice architecture and also enable and ease some of the presented real-world use-cases and prototypes.

As mentioned in section \ref{easlitArch} we are using different data stores for different use-cases. Our current data store lacks the native ability for version management (of items). Thus we plan to incorporate the quit-store, a graph store which is based on the distributed version management software Git, presented by Arndt et al. in \cite{arndt-n-2018--jws}. This new store fits natively into the microservice architecture and may be used whenever end users need access to former data versions. Furthermore it allows efficient exchange of data via the git distributed repository strategy, usable for data sharing between different EAs.LiT instances, e.g. hosted by different universities.

\section{Supplemental Online Material}

The application EAs.LiT consists of several services, licensed under Mozilla Public License Version 2. The source code of these services is available as different Gitlab repositories, listed as of this filtered project overview of the tech4comp Gitlab group at \url{https://gitlab.com/Tech4Comp?filter=eas}.

\bibliographystyle{splncs04}
\bibliography{bibliography}
\end{document}